# LOW POWER-AREA DESIGNS OF 1BIT FULL ADDER IN CADENCE VIRTUOSO PLATFORM

Karthik Reddy. G

Department of Electronics and Communication Engineering,
G. Pulla Reddy Engineering college, Kurnool, A.P, India
`karthik.reddy401@gmail.com`

*ABSTRACT*

*Power consumption has emerged as a primary design constraint for integrated circuits (ICs). In the Nano meter technology regime, leakage power has become a major component of total power. Full adder is the basic functional unit of an ALU. The power consumption of a processor is lowered by lowering the power consumption of an ALU, and the power consumption of an ALU can be lowered by lowering the power consumption of Full adder. So the full adder designs with low power characteristics are becoming more popular these days. This proposed work illustrates the design of the low-power less transistor full adder designs using cadence tool and virtuoso platform, the entire simulations have been done on 180nm single n-well CMOS bulk technology, in virtuoso platform of cadence tool with the supply voltage 1.8V and frequency of 100MHz. These circuits consume less power with maximum (6T design)of 93.1% power saving compare to conventional 28T design and 80.2% power saving compare to SERF design without much delay degradation. The proposed circuit exploits the advantage of GDI technique and pass transistor logic.*

*KEYWORDS*

*leakage power, GDI, Pass transistor logic, tri-state inverters.*

## 1. INTRODUCTION

Full adder circuit is functional building block and most critical component of complex arithmetic circuits like microprocessors, digital signal processors or any ALUs. Almost every complex computational circuit requires full adder circuitry. The entire computational block power consumption can be reduced by implementing low power techniques on full adder circuitry.

Several full adder circuits have been proposed targeting on design accents such as power, delay and area. Among those designs with less transistor count using pass transistor logic have been widely used to reduce power consumption [2-4]. In spite of the circuit simplicity, these designs suffer from severe output signal degradation and cannot sustain low voltage operations [5].

In these proposed designs we have exploited the advantages of GDI technique and PTL technique for low power. In these designs, we have generated carry using GDI technique (12T design Fig 9), we have generated carry using PMOS and NMOS pass transistors (8T design Fig 11) and also by using modified multiplexer using pass transistors (10T design Fig 10). The motivation is to use the tri-state inverter instead of inverter as it reduces power consumption by 80% when compare to normal inverter. And sum is generated using 6T XOR module as shown in Fig.7.

In these designs we have exploited the advantages of GDI technique and PTL technique for low power. In these designs, we have generated carry using GDI technique, we have generated carry using PMOS and NMOS pass transistors and also by using modified multiplexer using pass transistors. The motivation is to use the tri-state inverter instead of inverter as it reduces power consumption by 80% when compare to normal inverter. And sum is generated using 6T XOR module as shown in Fig.7.

The rest of the paper is organised as previous research work, proposed full adder designs, simulations-results-comparison and conclusion.

## 2. PREVIOUS WORK

Many full adder designs have been reported using static and dynamic styles in papers [1-4]. These designs can be divided into two types, the CMOS logic and the pass-transistor logic [5]. Different full adder topologies have been proposed using standard XOR and XNOR circuits and with 3T XOR-XNOR modules.

In [5] a low power full adder cell has been proposed, each of its XOR and XNOR gates has 3 transistors. Advantages of pass-transistor logic and domino logic encouraged researchers to design full adder cell using these concepts [6] [7]. Full adder cells based on Sense energy recovery full adder (SERF) [8] and Gate diffusion input (GDI) techniques [5] are common. To attain low power and high speed in full adder circuits, pseudo-NMOS style with inverters has been used [9]. A 10 transistors full adder using top-down approach [10] and hybrid full adder [11] are the other structures of full adder cells. Sub threshold 1-Bit full adder cell and hybrid CMOS design style are the other techniques that targeted on fast carry generation and low PDP.

Many PTL circuit implementations have been proposed in the earlier papers [6], [7]. Some of the main advantages of PTL over standard CMOS design are 1) high speed, due to the small node capacitances; 2) low power dissipation, as a result of the reduced number of transistors; and 3) lower interconnection effects [7], [6], due to a small area. However, most of the PTL implementations have two basic problems. Firstly, the threshold drop across the single-channel pass transistors results in reduced current drive and hence slower operation at reduced supply voltages; this is particularly important for low-power design since it is desirable to operate at the lowest possible voltage level. Secondly, since the "high" input voltage level at the regenerative inverters is not, the PMOS device in the inverter is not fully turned off, and hence direct-path static power dissipation could be significant [4]. In this paper 3 new designs of full adder circuits have been proposed.

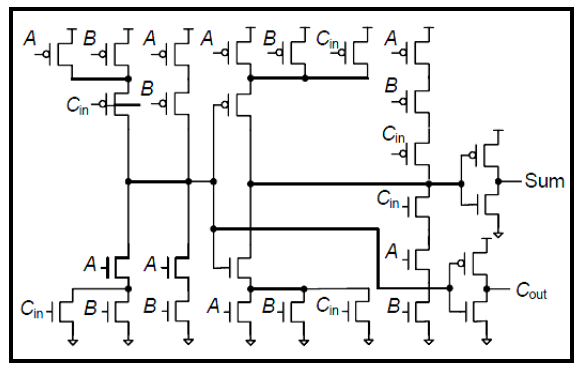

Fig.1. conventional 28T full adder

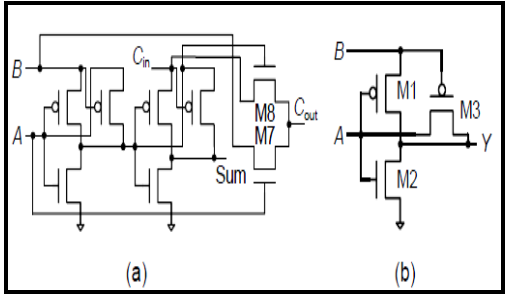 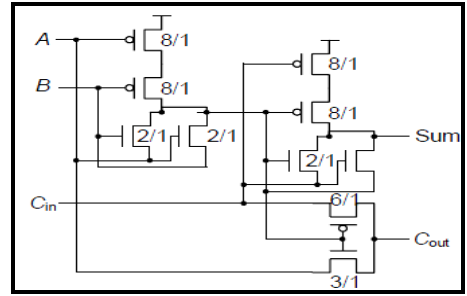

Fig.2. Design of chowdhury *etal.(2008)*          Fig.3. SERF full adder design

(a).8T full adder, (b) 3T XOr gate

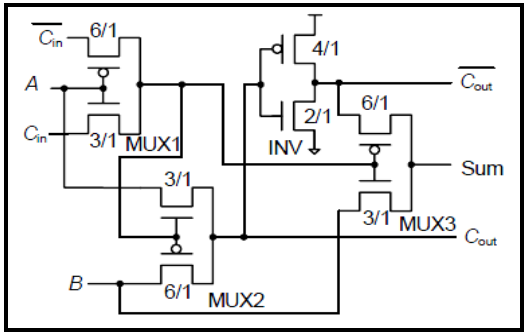 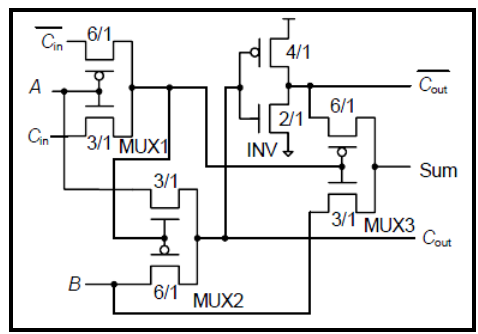

Fig.4. 8T full adder design [15]          Fig.5 8T full adder design [16]

## 3. DESIGN OF PROPOSED FULL ADDER CIRCUITS

### 3.1. 3T XOR gate and tri-state inverter design

Most full adder designs with less transistor count adopt 3-module implementations i.e. XOR (or XNOR), for sum as well as carry modules [1]. For PTL based designs, it requires at least 4 transistors to implement a XOR (or XNOR) module [5, 8] but the design faces severe threshold voltage loss problems.

The motivation for these designs is use of tri-state inverter instead of normal inverter because tri-state inverter's power consumption is 80% less than normal inverter. In normal inverter the supply voltage is always HIGH; while in the tri-state inverter the supply voltage is not always HIGH. This reduces the average leakage of the circuit throughout operation. The diagram for tri-state inverter is shown on Fig. 6.

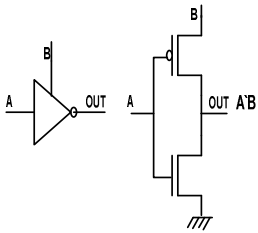 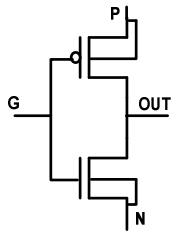 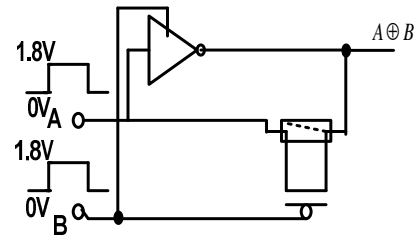

Fig.6. Tristate inverter          Fig.8. Basic GDI cell          Fig.7. 3T XOR module

Note: Switching of the MOS transistor is also shown in fig. 7 and it is repeated in all figures

### 3.2. Proposed 12T full adder design

The proposed 12T full adder design incorporates the 3T XOR module made by tri-state inverter as shown in Fig.7. The design follows with the conventional 2 module implementation of 3 input XOR gate, this facilitate sum module of the full adder.

The modified equations (1) for 12T full adder design are:

$$\begin{aligned} sum &= a \oplus b \oplus c \\ &\Rightarrow (a \oplus b) \oplus c \\ carry &= ab + bc + ca \\ ab + bc + ca &= ab + bc(a + a`) + ac(b + b`) \\ &\Rightarrow ab + abc + a`bc + ab`c \\ &\Rightarrow ab(1 + c) + (a`b + ab`)c \\ &\Rightarrow ab + (a \oplus b)c \end{aligned} \quad (1)$$

The sum is generated by implementing 3T XOR module twice. Carry module is generated here by using GDI technique.

The GDI approach allows implementation of a wide range of complex logic functions using only two transistors. This method is suitable for design of fast, low-power circuits, using a reduced number of transistors (as compared to CMOS and existing PTL techniques), while improving logic level swing and static power characteristics.

1) The GDI cell contains three inputs G(common gate input of nMOS and pMOS), P (input to the source/drain of pMOS), and N (input to the source/drain of nMOS).
2) Body of both nMOS and pMOS are connected to N or P (respectively) as shown in Fig.8. , so it can be arbitrarily biased at contrast with a CMOS inverter.

This circuit exploits the low power advantages of GDI circuits to generate carry and tri-state inverter for generating sum. The equations have modified as above to generate carry.

Basic operations like AND, OR have performed using GDI technique to generate carry, for example in the equation (1) $a.b$ and $(a \oplus b)c$ have been performed using GDI and gates. Sum is implemented by using 3T (XOR) module twice as shown in Fig.9.

### 3.3. Proposed 10T full adder design

The proposed 10T full adder uses the concept of pass transistor logic based multiplexer. The pass transistor design reduces the parasitic capacitances and results in fast circuits. The multiplexer is implemented using pass transistors for carry generation. This design is simple and efficient in terms of area and timing. The proposed 10T full adder circuit can be visualised by modifying the equations (2) as accordingly

The modified equations for 10T full adder design:

$$sum = a \oplus b \oplus c$$
$$\Rightarrow (a \oplus b) \oplus c$$
$$carry = ab + bc + ca \qquad (2)$$
$$\Rightarrow ab + (a \oplus b)c$$
$$\Rightarrow ab(a \oplus b)` + (a \oplus b)c$$

The multiplexer using pass transistor logic can be visualised in 2T model, the select signal for the multiplexer here is ($a \oplus b$). The equations have modified such that select signal is in the form of ($a \oplus b$). The ($a.b$) signal is generated by using the tri-state inverter for

$$(a`b) => (a`b)`b = (a+b`).b \Rightarrow a.b + b.b` \Rightarrow a.b. \qquad (3)$$

The sum is generated by implementing 3T XOR module twice. Carry is generated by using pass transistor logic based multiplexer whose select line is ($a \oplus b$) as shown in Fig.10.

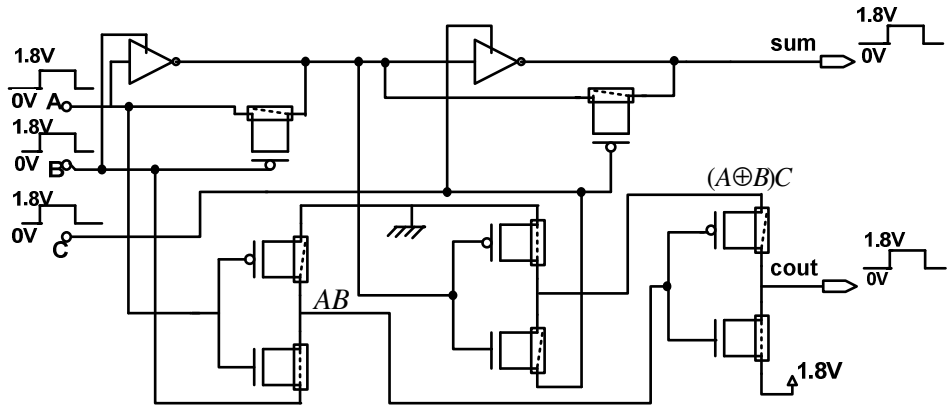

Fig.9. proposed 12T full adder design

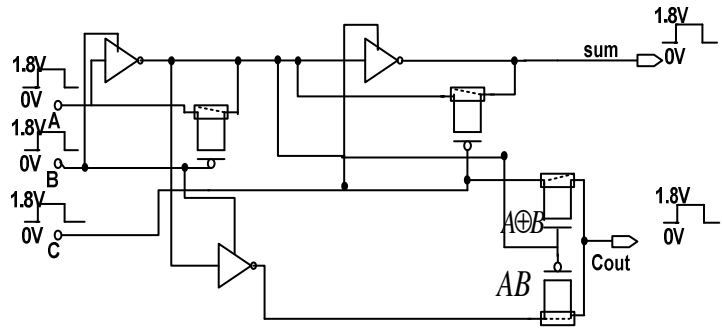

Fig.10. proposed 10T full adder design

### 3.4. Proposed 8T full adder design

In the proposed 8T full adder sum is generated using 3T XOR module twice, and carry is generated using NMOS and PMOS pass transistor logic devices as shown in Fig.11. The equations (4) are modified so as to visualise the 8T full adder design.

The modified equations for 8T full adder design are:

$$sum = a \oplus b \oplus c$$
$$\Rightarrow (a \oplus b) \oplus c$$
$$carry = ab + bc + ca$$
$$\Rightarrow ab + bc(a + a`) + ac(b + b`)$$
$$\Rightarrow ab + (a \oplus b)c$$
$$\Rightarrow (a`b)`b + (a \oplus b)c$$

(4)

In this design instead of using two NMOS pass transistor devices we have used one NMOS and one PMOS pass transistor device, because of ease of the design and as according to the equation as shown in Fig.11.

It must be noted that PMOS transistor passes ' 1' very good, but cannot pass '0' completely thus, the carry output has weak '0'. NMOS transistor passes '0' very good, but cannot pass '1' completely therefore, the carry output has weak ' 1 ', Having weak '0' and '1' at carry outputs is one of the disadvantages of proposed 8T full adder circuit. In practical situations, this problem can be solved by using an inverter at carry output, but this solution leads to increased power and area.

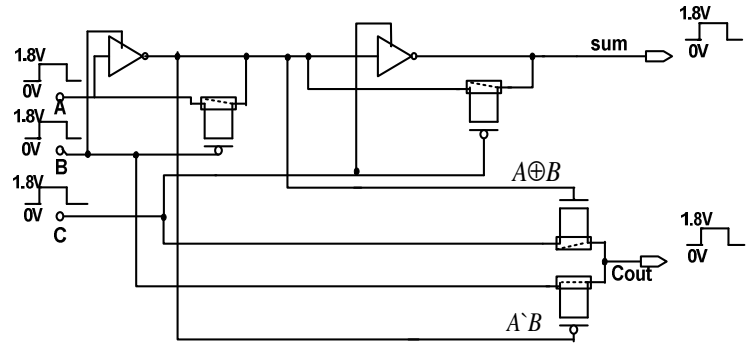

Fig.11. Proposed 8T full adder design

### 3.5 Proposed 6T full adder design

In the proposed 6T full adder sum is generated using 2T XOR module twice, and carry is generated using NMOS and PMOS pass transistor logic devices as shown in Fig.12. The equations (Eq 5) are modified so as to visualise the 6T full adder design.

The modified equations for 6T full adder design are:

$$sum = (a \oplus b) \oplus c$$
$$\Rightarrow (a \oplus b)c` + (a \oplus b)`c$$
$$carry = ab + bc + ca$$
$$\Rightarrow (a \oplus b)`a + (a \oplus b)c$$

(5)

In this design (a xor b) signal is passed to the pass transistor multiplexer made of two transistors to choose one among two. To generate carry (a xor b) is sent to multiplexer to choose between a, c. and to generate sum (a xor a) is sent to choose between c`, c.

The entire simulations for all Full adders have been done on 180nm, single n-well CMOS bulk technology, in virtuoso platform of cadence tool with the supply voltage 1.8V and frequency of 100MHz. The entire results are compared with the different full adder designs. Area is calculated by using MICROWIND software. The area is reduced by 48% for proposed 12T design, the area is reduced by 66% for proposed 8T design and area is reduced by 53% when compared to 28T conventional full adder design.

The simulation results show that the power dissipation is very less compared to any full adder design. Hence this design is used in ALU design.

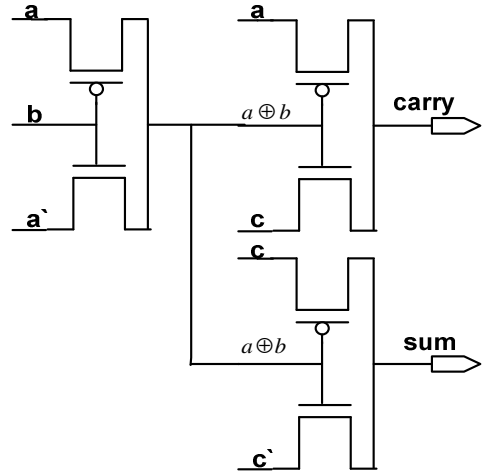

Fig.12. Proposed 6T full adder design

## 4. SIMULATION RESULTS AND COMPARISONS

The entire simulations have been done on 180nm single n-well CMOS bulk technology, in virtuoso platform of cadence tool with the supply voltage 1.8V and frequency of 100MHz. The entire results are compared with the different techniques. Area is calculated by using micro wind software. The area is reduced by 48% for proposed 12T design, the area is reduced by 66% for proposed 8T design and area is reduced by 53% when compared to 28T conventional full adder design.

Table 1. Simulation results.

| Full Adder Designs | Conventional (28T) | Chowdury deign(8T) | SERF ref[55] | Proposed (12T) | Proposed (8T) | Proposed (10T) | Proposed(6T) |
|---|---|---|---|---|---|---|---|
| Cout delay (nS) | 0.366 | 0.513 | 0.39 | 0.476 | 0.502 | 0.512 | 0.36 |
| Avg.power consumption (uW) | 52.4 | 17.4 | 18.2 | 14.3 | 16.0 | 17.1 | 3.6 |
| Number-of transistors | 28 | 8 | 10 | 12 | 8 | 10 | 6 |
| Power*Delay (uW.nS) | 19.178 | 8.926 | 7.09 | 6.806 | 8.032 | 8.755 | 1.296 |

## 5. CONCLUSIONS

Four new full adder designs have been proposed and simulation results have been compared with the previous results in UMC180nm technology using cadence tool. According to the simulation results these circuits consume less power with maximum (6T design)of 93.1% power saving compare to conventional 28T design and 80.2% power saving compare to SERF design without much delay degradation.

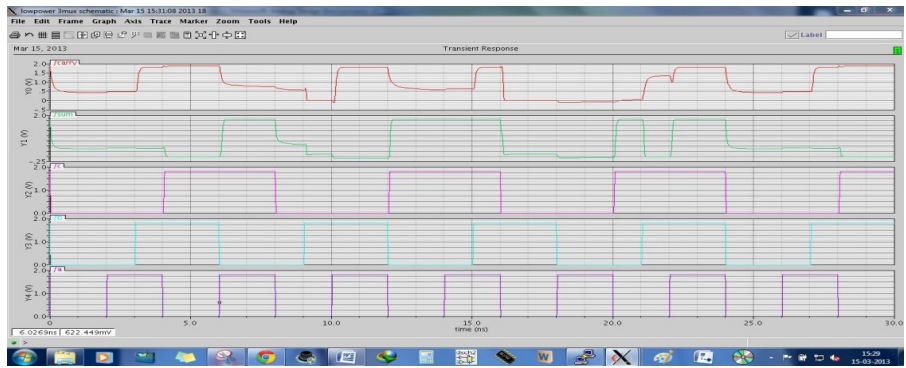

Fig. 13 Transient response of 12T full adder design

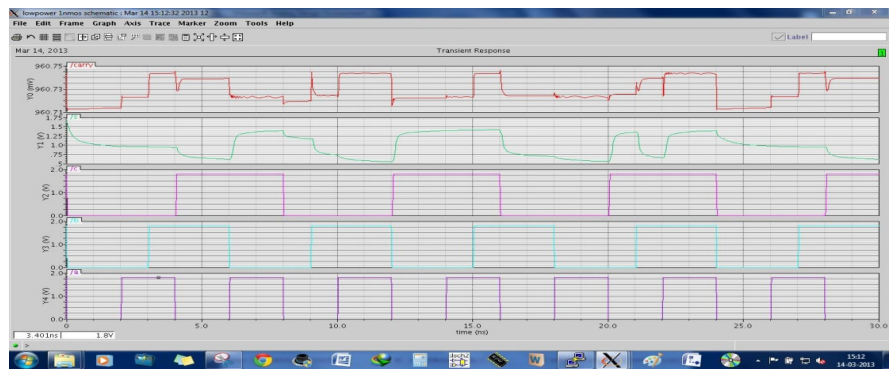

Fig.14 Transient response of 10T full adder design

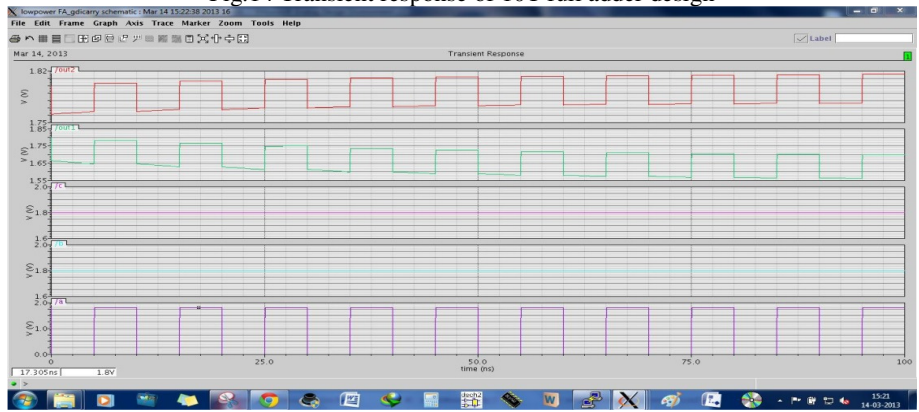

Fig.15 Transient response of 8T full adder design

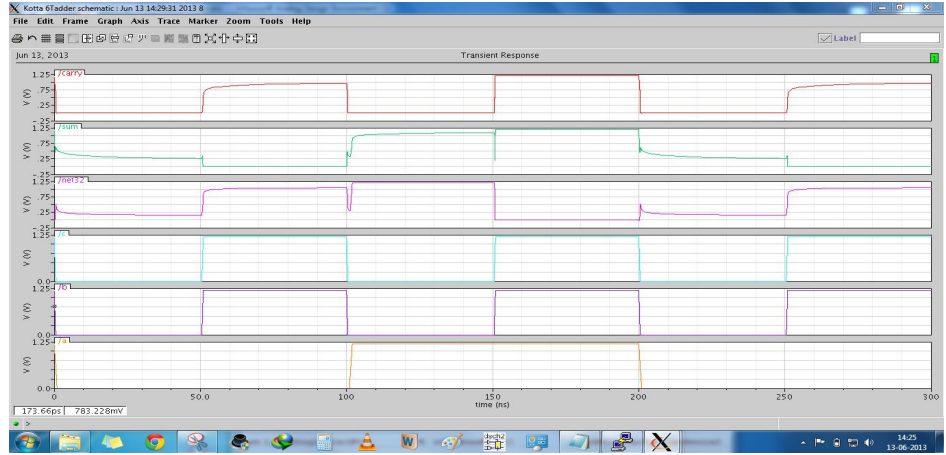

Fig.16 Transient response of 6T full adder design

## REFERENCES

[1] A. Fayed and M. A. Bayoumi, "A low-power 10 transistor full adder cell for embedded architectures," in Proc. IEEE Int. Symp. Circuits Syst., 2001, pp.226–229.
[2] H. T. Bui, Y. Wang, and Y. Jiang, "Design and analysis of low-power 10-transistor full adders using XOR XNOR gates," IEEE Trans. Circuits Syst. II, Analog and Digital Signal Processing., vol.49, no. 1, pp. 25–30, Jan. 2002.
[3] J.-F. Lin, Y.-T. Hwang, M.-H. Sheu and C.-C. Ho, "A novel high speed and energy efficient 10-transistor full adder design," IEEE Trans. Circuits Syst. I, vol. 54, no. 5, pp. 1050–1059, May 2007.
[4] Y. Jiang, Al-Sheraidah. A, Y. Wang, Sha. E, and J. G. Chung, "A novel multiplexer-based low-power full adder," IEEE Trans. Circuits Syst. II, Analog Digit. Signal Process., vol. 51, pp.345–348, July 2004.
[5] Dan Wang, Maofeng Yang, Wu Cheng XUguang Guan, Zhangming Zhu, Yintang Yang " Novel Low power Full Adder Cells in 180nm CMOS Technology", 4th IEEE conference on Industrial Electronics and Applications, pp. 430-433,2009.
[6] Sreehari Veeramachaneni, M.B. Srinivas, "New improved I•bit full adder cells ", Canadian Conference on Electrical and Computer Engineering, pp. 000735 -000738, 2008.
[7] Chuen•Yau, Chen and, Yung•Pei Chou, "Novel Low•Power I•bit Full Adder Design", 9th International Symposium on Communications and Information Technology, pp. 1348 -1349, 2009
[8] F.Moradi, DTWisland, H.Mahmoodi, S.Aunet; T.V.Cao, A.Peiravi, "Ultra low power full adder topologies", IEEE International Symposium on Circuits and Systems, pp. 3158-3161, 2009.
[9] Amir Ali Khatibzadeh, Kaamran Raahemifar, "A 14•TRANSISTOR LOW POWER HIGH•SPEED FULL ADDER CELL", Canadian Conference on Electrical and Computer Engineering, vol. I, pp. 163-166, 2003.
[10] AK. Singh, C.M.R. Prabhu, K.M.Almadhagi, S.F. Farea, K. Shaban,"A Proposed 10•T Full Adder Cell for Low Power Consumption", International Conference on Electrical Engineering/Electronics Computer Telecommunications and Information Technology (ECTI•CON), pp. 389 -391, 2010.
[11] IIham Hassoune, Denis Flandre, .Iean•Didier Legat," ULPF A: A New Efficient Design of a Power• Aware Full Adder", IEEE Transactions on Circuits and Systems I: Regular Papers, Vol. 57, pp. 2066 - 2074, August 2010.
[12] S. Goel, A. Kumar, and M. A. Bayoumi, "Design of robust, energy-efficient full adders for deep submicrometer design using hybrid-CMOS logic style," IEEE Trans. Very Large Scale Integr. (VLSI) Syst., vol. 14, no. 12, pp.1309–1321, Dec. 2006.
[13] H. T. Bui, Y. Wang, and Y. Jiang, "Design and analysis of low-power 10-transistor full adders using XOR–XNOR gates," IEEE Trans. Circuits Syst. II, Analog Digit. Signal Process., vol. 49, no. 1, pp. 25–30, Jan. 2002.
[14] J.-F. Lin, Y.-T. Hwang, M.-H. Sheu and C.-C. Ho, "A novel high speed and energy efficient 10-transistor full adder design," IEEE Trans. Circuits Syst. I, vol. 54, no. 5, pp. 1050–1059, May 2007.

[15] Yi WEI, Ji-zhong SHEN, "Design of a novel low power 8-transistor 1-bit full adder cell", Journal of Zhejiang University-SCIENCE C (Computers & Electronics), vol. 7,pp 504-507, Dec 2011.

[16] Nabiallah Shiri Asmangerdi, Javad Forounchi and Kuresh Ghanbari, "A new 8- Transistor Floating Full-Adder Circuit", IEEE Trans. 20th Iranian Conference on Electrical Engineering, (ICEE2012), pp. 1405-1409, May, 2012.

**Author Biography**

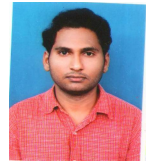

G. Karthik Reddy completed his Bachelor of technology in ECE branch in Mahatma Gandhi Institute of Technology, Hyderabad, India in 2010. He completed his Master of technology in VLSI & Embedded systems specialization at Maulana Azad National Institute of Technology, Bhopal, India in 2013. He is working as Assistant Professor in ECE department at G. Pulla Reddy Engineering college, Kurnool, India, his area of interest include Low Power VLSI design.